\documentclass[lettersize,journal]{IEEEtran}
\usepackage{enumitem}
\usepackage{amsmath,amsfonts}
\usepackage{algorithmic}
\usepackage{algorithm}
\usepackage{array}
\usepackage[caption=false,font=normalsize,labelfont=sf,textfont=sf]{subfig}
\usepackage{textcomp}
\usepackage{stfloats}
\usepackage{verbatim}
\usepackage{graphicx}
\usepackage[noadjust]{cite}
\usepackage[dvipsnames,table]{xcolor}
\usepackage{tikz}
\usepackage{pgfplots}
\pgfplotsset{compat=1.16}
\usepackage{amsmath}
\DeclareMathOperator*{\minimize}{minimize}
\DeclareMathOperator*{\maximize}{maximize}
\usepackage{gensymb}
\usepackage{environ}
\makeatletter
\newsavebox{\measure@tikzpicture}
\NewEnviron{scaletikzpicturetowidth}[1]{%
  \def\tikz@width{#1}%
  \def\tikzscale{1}\begin{lrbox}{\measure@tikzpicture}%
  \BODY
  \end{lrbox}%
  \pgfmathparse{#1/\wd\measure@tikzpicture}%
  \edef\tikzscale{\pgfmathresult}%
  \BODY
}
\def\checkmark{\tikz\fill[scale=0.4](0,.35) -- (.25,0) -- (1,.7) -- (.25,.15) -- cycle;}
\makeatother
\allowdisplaybreaks
\usepackage[hyphens]{url}
\usepackage{hyperref} 

\usepackage{makecell}
\usepackage{hhline}
\usepackage{ulem}

\DeclareUnicodeCharacter{2212}{-}

\DeclareMathOperator{\CVaR}{CVaR}

\DeclareMathOperator{\VaR}{VaR}

\begin{document}

\title{Reserve Provision from Electric Vehicles: Aggregate Boundaries and Stochastic Model Predictive Control}
\author{Jacob Thrän, \IEEEmembership{Student Member, IEEE}, Jakub Mareček, \IEEEmembership{Member, IEEE}, Robert N. Shorten, Timothy C. Green, \IEEEmembership{Fellow, IEEE}
\thanks{Jacob Thrän, Robert N. Shorten, and Timothy C. Green are with Imperial College, London, UK.}
\thanks{Jakub Mareček is with the Czech Technical University, Prague, Czechia.}} 
\IEEEaftertitletext{\vspace{-2.5\baselineskip}}
\markboth{IEEE Transactions on Power Systems,~Vol.~X, No.~X, Month X~20XX}%
{Shell \MakeLowercase{\textit{et al.}}: A Sample Article Using IEEEtran.cls for IEEE Journals}

\IEEEpubid{0000--0000/00\$00.00~\copyright~2021 IEEE}

\onecolumn
\thispagestyle{empty}\setcounter{page}{0}%
\vfill
{\centering\Huge Reserve Provision from Electric Vehicles: Aggregate
Boundaries and Stochastic Model Predictive Control \par}

\vspace{5mm}

\noindent{Jacob Thrän; Jakub Mareček; Robert N. Shorten; Timothy C. Green}

\vspace{15mm}

\noindent Document version:

\noindent Accepted peer-reviewed manuscript

\vspace{10mm}

\noindent \textbf{This paper has been accepted by IEEE Transactions on Power Systems, with its copyright
transferred to IEEE:}

\noindent J. Thrän, J. Mareček, R. Shorten, T.C. Green, “Reserve Provision from Electric Vehicles: Aggregate Boundaries and Stochastic Model Predictive Control”, in IEEE Transactions on Power Systems, doi: 10.1109/TPWRS.2025.3539863

\vspace{5mm}

\noindent \textcolor{blue}{\underline{\url{https://ieeexplore.ieee.org/document/10878458}}}

\vfill
\twocolumn

\maketitle

\begin{abstract}
Controlled charging of electric vehicles, EVs, is a major potential source of flexibility to facilitate the integration of variable renewable energy and reduce the need for stationary energy storage. To offer system services from EVs, fleet aggregators must address the uncertainty of individual driving and charging behaviour. This paper introduces a means of forecasting the service volume available from EVs by considering several EV batteries as one conceptual battery with aggregate power and energy boundaries. Aggregation avoids the difficult prediction of individual driving behaviour. The predictability of the boundaries is demonstrated using a multiple linear regression model which achieves a normalised root mean square error of 20\% - 40\% for a fleet of 1,000 EVs. A two-stage stochastic model predictive control algorithm is used to schedule reserve services on a day-ahead basis addressing risk trade-offs by including Conditional Value-at-Risk in the objective function. A case study with 1.2 million domestic EV charge records from Great Britain illustrates that increasing fleet size improves prediction accuracy, thereby increasing reserve revenues and decreasing an aggregator's operational costs. For fleet sizes of 400 or above, cost reductions plateau at 60\% compared to uncontrolled charging, with an average of 1.8kW of reserve provided per vehicle.
\end{abstract}


\iftrue
\vspace{-3mm}
\section*{Nomenclature}
\addcontentsline{toc}{section}{Nomenclature}
\noindent \textit{Abbreviations}
\begin{IEEEdescription}[\IEEEusemathlabelsep\IEEEsetlabelwidth{$b^{PR}_t/b^{NR}_t$}]
\item[V2G] Vehicle-to-grid
\item[G2V] Grid-to-vehicle
\item[SOC] State of Charge
\item[MLR] Multiple Linear Regression
\item[CVaR] Conditional Value-at-Risk
\item[VaR] Value-at-Risk
\item[SMPC] Stochastic Model Predictive Control
\item[MILP] Mixed-Integer Linear Programming
\end{IEEEdescription}
\textit{Indices/Sets}
\begin{IEEEdescription}[\IEEEusemathlabelsep\IEEEsetlabelwidth{$b^{PR}_t/b^{NR}_t$}]
\item[$i$/$I$] Index/Set of EVs in an EV fleet
\item[$t$/$T$] Index/Set of settlements in a planning window
\item[$s$/$S$] Index/Set of scenarios
\item[$j$/$J$] Index/Set of charging events at a single charger
\end{IEEEdescription}
\textit{Parameters}\IEEEpubidadjcol
\begin{IEEEdescription}[\IEEEusemathlabelsep\IEEEsetlabelwidth{$b^{PR}_t/b^{NR}_t$}]
\item[$\Delta t$] Settlement period duration
\item[$\Delta t_r$] Maximum duration of reserve activation
\item[$t_{auc}$] Settlement period of reserve auction
\item[$t_{x}$] Current settlement period for energy trajectory
\item[$t^a_i$/$t^d_i$] Arrival/departure time of EV $i$
\item[$\eta$] Charging efficiency
\item[$p^{max}$] Rated charging power in $kW$
\item[$e^{a}$] Energy level in EV battery on arrival in $kWh$
\item[$e^{max}$] Maximum energy capacity of EV battery in $kWh$
\item[$e^{d}$] Expected energy in EV battery upon departure in $kWh$
\item[$e^{min}$] Minimum allowed energy level in EV battery in $kWh$
\item[$t_i^a$/$t_i^d$] Plug-in/plug-out time of EV $i$
\item[$E^u_t$/$E^l_t$] Upper/lower aggregate energy boundary in $kWh$
\item[$P^b_t$] Aggregate Power Boundary in $kW$
\item[$E^d_t$] Difference between energy boundaries in $kWh$
\item[$E^{EV}_t$] Aggregate charging trajectory in $kWh$
\item[$c^{PR}_t$/$c^{NR}_t$] Reward for providing positive/negative reserve in $\pounds/kW$
\item[$c^E_t$] Cost of charging in $\pounds/kWh$
\item[$\bar{c}^E$] Average cost of charging in $\pounds/kWh$
\item[$z$] Number of settlements considered in baselining
\item[$M_1,M_2$] Large constants to ensure constraints are non-limiting
\item[$A_s$] Compensation for end-of-plan energy in $\pounds$
\item[$p_s$] Probability of scenario $s$
\item[$\Omega$] Risk aversion factor
\item[$\alpha$] CVaR probability threshold
\end{IEEEdescription}
\textit{SMPC Variables}
\begin{IEEEdescription}[\IEEEusemathlabelsep\IEEEsetlabelwidth{$b^{PR}_t/b^{NR}_t$}]
\item[$P^{NR,g}_t$] Negative reserve power on grid side in $kW$
\item[$P^{PR,g}_t$] Positive reserve power on grid side in $kW$
\item[$P^{NR,v}_t$] Negative reserve power on vehicle side in $kW$
\item[$P^{PR,v}_t$] Positive reserve power on vehicle side in $kW$
\item[$P^{V2G,g}_t$] Vehicle-to-grid power in $kW$
\item[$P^{G2V,g}_t$] Grid-to-vehicle power in $kW$
\item[$\Delta P^{pen}$] Non-delivered power in $kW$
\item[$C^{cha}_s$] Charging costs in scenario $s$ in $\pounds$
\item[$C^{pen}_s$] Penalty payment in scenario $s$ in $\pounds$
\item[$R^{res}$] Revenue from reserve provision in $\pounds$
\item[$b^{PR}_t/b^{NR}_t$] Binary reserve commitment variable 
\item[$w_s$] Auxiliary variable to derive CVaR from VaR
\end{IEEEdescription}
\textit{MLR Variables}
\begin{IEEEdescription}[\IEEEusemathlabelsep\IEEEsetlabelwidth{$b^{PR}_t/b^{NR}_t$}]
\item[$\beta_{m,t}$] Regression coefficients
\item[$H_t$] Temperature at time $t$ in $\degree C$
\item[$W_t$] Precipitation at time $t$ in $mm$
\item[$B_t$] Binary variable for bank holidays
\item[$D_t$] Categorical variable for the day of the week
\item[$\varepsilon_t$] Random regression error
\end{IEEEdescription}
\fi

\IEEEpubidadjcol

\section{Introduction}
Electric vehicles (EVs) are set to play a key role in the energy transition. Not only are they potentially a zero-carbon option for individual transport, but their batteries also offer the possibility of providing flexibility services to a decarbonised grid. It is predicted that by 2030, the battery capacity installed in EVs will be more than eight times the global capacity of stationary battery storage \cite{staffellIAEE}. With the rise of variable renewable energy (VRE) leading to large investments in energy storage, the use of EV battery capacity for grid services could reduce system costs and provide revenue to consumers. The benefits of so-called demand response (DR) from consumer-owned equipment have been laid out clearly \cite{strbac2008demand} but the uptake of DR, so far, has disappointed previous predictions \cite{https://doi.org/10.1002/er.3683}. The global rise of EVs now presents new possibilities for DR because these are large loads and can be manipulated with little or no interference with consumers' lives. Further, the possibility of feeding power back from vehicle-to-grid (V2G) yields a larger volume of DR. Consumer behaviour, however, remains inherently uncertain, and this complicates the introduction of grid services, such as frequency regulation or reserve, from EV charging, because these services require reliability. Predicting consumer behaviour at an individual level is very difficult \cite{9797708}, with prediction algorithms struggling to achieve normalised root mean square errors below 45\% for individual EV energy consumption forecasts \cite{shahriar2021prediction}. Aggregate consumer behaviour, on the other hand, can be predicted with some level of certainty, even when there is little knowledge about individual behaviour \cite{travacca2017stochastic}. This leads to considering aggregate measures of the available flexibility of an EV fleet which would facilitate both ancillary service scheduling and energy arbitrage so that consumers earn revenue for the former and reduce costs through the latter. It remains important to provide grid services while causing the least possible disruption to the consumer's everyday life \cite{crisostomi2017electric}.

A number of deterministic models have been proposed to create aggregate measures of flexible charging characteristics \cite{7410071,7463483,xu2020greenhouse,yu2015balancing,pertl2018equivalent} with the stated objective of computational efficiency. \cite{wang2022integrating} compares how well these models can be integrated into wider energy system models where computational efficiency is at a premium. None of the compared studies investigate how their aggregate models can be used to create accurate stochastic predictions for EV scheduling. This may be because of their focus on energy system modelling rather than charge scheduling. Instead, a separate body of literature considers stochastic scheduling algorithms for individual EVs to deal with uncertain driving and charging behaviour \cite{nimalsiri2019survey}. \cite{tian2020risk} proposed an algorithm based on downside risk constraints (DRC) and \cite{aliasghari2020risk} used a hybrid approach between stochastic programming and information gap decision theory (IGDT). \cite{wang2020scenario} and \cite{gong2022model} proposed stochastic optimisation approaches with the first considering V2G charging whereas the latter only studied uni-directional flexible charging. \cite{han2019day} and \cite{8274706} considered uncertain future energy and service prices stochastically but did not treat driving behaviour as uncertain. Stochastic programming has also been used to schedule electric bus charging \cite{9852789} and to schedule EV charging in the context of a virtual power plant with a range of distributed energy resources \cite{https://doi.org/10.1049/enc2.12018}. All of the listed stochastic approaches considered charging uncertainty on an individual EV level, though, and none of them took advantage of an aggregation model to manage the inherent uncertainty of charging behaviour.

\setlength{\tabcolsep}{3.5pt}
\begin{table}[t]

    \centering
        \caption{Comparison to recent studies}
    \label{tab:litreview}
    \resizebox{\linewidth}{!}{%
    \begin{tabular}{l | c c c | c c}
        \hline
        \hline
         Study & \makecell[cl]{Aggregate\\Model} & \makecell[cl]{Stochastic\\Scheduling} & \makecell[cl]{Data-Based\\Predictions} & \makecell[cl]{Ancillary\\Services} & CVaR\\
         \hline
         \cite{7410071} & \checkmark & & & & \\
         \cite{7463483,xu2020greenhouse,yu2015balancing} & \checkmark & & & \checkmark & \\
         \cite{pertl2018equivalent} & \checkmark & & \checkmark & & \\
         \cite{tian2020risk,han2019day,8274706,9852789} & & \checkmark & & \checkmark  & \\
         \cite{aliasghari2020risk} & & \checkmark & \checkmark & & \\
         
         \cite{wang2020scenario} & & \checkmark & \checkmark & \checkmark &  \\

         \cite{gong2022model} & & \checkmark & & \checkmark & \checkmark\\
         \cite{https://doi.org/10.1049/enc2.12018} & & \checkmark & & & \\
         \cite{10415236} & \checkmark & \checkmark & & & \checkmark \\
         \hline
         This study & \checkmark & \checkmark & \checkmark & \checkmark & \checkmark\\
         \hline\hline
        
    \end{tabular}}

\end{table}

Combining aggregation models, stochastic algorithms, and data-based predictions is important when realistically considering EV scheduling without future knowledge. The aggregate models allow the summation of a range of complex characteristics into a few aggregate variables, which can then be forecast using prediction models. Aggregation increases the accuracy of predictions because aggregated variables follow the law of large numbers (which states that the average of a sample converges to the true value as sample size increases). For an EV fleet, this means that an aggregate measure, e.g. the percentage of plugged-in EVs, converges as the number of vehicles increases. To account for the remaining error in the predictions, stochastic programming is essential, as the charging algorithm would otherwise be incapable of preparing for even the slightest deviation from the predictions. Lastly, using actual data-based predictions, rather than assuming future knowledge or using assumed scenario distributions, is important to give a realistic image of the algorithm’s performance and offers valuable insights into the effect of fleet size on financial performance, permitting the observation of any thresholds for profitable operation. Table~\ref{tab:litreview} summarises recent papers that have considered at least one of those three aspects. \cite{10415236} added stochasticity to an aggregate model (\cite{7463483}), and \cite{pertl2018equivalent} tested the predictability of the measures from their own aggregate model but there are, to the best of the authors' knowledge, no studies that stochastically forecast the charging characteristics from any of the aggregation models \cite{7463483,xu2020greenhouse,yu2015balancing,pertl2018equivalent} based on real charging data. This gap is addressed in this paper by combining stochastic scheduling with the aggregation model from \cite{7463483}, and building a prediction model for its boundaries. This creates an intricate multivariate stochastic optimisation which, for the first time, combines aggregation, stochasticity, and predictions. By putting the optimisation in a two-stage stochastic model predictive control (SMPC) framework, it can be applied to both real-time charge scheduling as well as day-ahead ancillary service auctions.

Table~\ref{tab:litreview} also states whether previous papers have considered ancillary service provision and whether they have included the risk measure Conditional Value-at-Risk (CVaR). Day-ahead ancillary service procurement carries the risk of prohibitively high penalties. Including a risk measure, like CVaR allows aggregators to manage this risk and assess the effect of risk aversion on financial performance. Previous CVaR implementations for EV scheduling have either failed to consider the risky day-ahead nature of ancillary services \cite{gong2022model}, or not considered ancillary service provision altogether \cite{10415236}. This paper is the first to demonstrate how CVaR can mitigate the risk of day-ahead ancillary service provision from V2G.

The model from \cite{7463483} was chosen over other aggregation models because its aggregate boundary approach remarkably allows all of the charging characteristics of an EV fleet to be summarised into only three variables. These result from the authors considering the batteries within an EV fleet to behave like one aggregate battery. Charging and discharging this virtual battery is constrained by a power boundary, while the battery capacity is constrained by an upper and a lower energy boundary. This reduction means that the model allows for NP-hard scheduling algorithms to be deployed regardless of EV fleet size. Furthermore, their proposed aggregate energy and power boundaries are well-suited to create aggregate predictions. \cite{7463483} [9] did not indicate, though, that their model requires energy transfers between different EVs within the virtual battery to function. For that reason, their aggregate boundaries are revisited in this paper, with important extensions to reflect the effect of battery cell voltage on available charging power. The detailed contributions of this paper are:
\begin{enumerate}[leftmargin=*]
    \item The aggregation procedure from \cite{7463483} is extended to explain the need for intra-EV-fleet energy transfers and account is taken of charge power variation with SOC.
    \item The proposed aggregate energy and power boundaries are tested for their forecastability with the introduction of a multiple linear regression (MLR) model. 
    \item Scenarios from the MLR model are combined with a two-stage stochastic optimisation to form an SMPC approach which combines ancillary service scheduling and real-time charge allocation. CVaR is included in the objective function as a risk measure to trade-off between additional service revenue and non-delivery penalties.
    \item The predictive and financial performance of the proposed algorithm is assessed in a case study on providing reserve services in the GB power grid, yielding an estimation of the EV fleet size required for the realistic provision of services.
\end{enumerate}

The remainder of the paper is structured as follows: Section~\ref{aggbound} revisits the aggregate boundaries, and section~\ref{smpcmodel} introduces the SMPC model including the MLR prediction method. Section~\ref{casestudy} demonstrates the new model in a case study on reserve services for the GB electricity system, the results of which are presented in section~\ref{results}.

\section{Energy and Power Boundaries of Aggregate Energy Storage} \label{aggbound}

  \cite{7463483} shows that the charging behaviour of an aggregate battery can be expressed through upper and lower energy boundaries and a power boundary. The maximal charge that can be added to the aggregate battery at a point in time is limited by the upper energy boundary which is set by the amount of energy that the connected EVs can accept. The minimal amount of energy that has to be present in the aggregate battery is the lower energy boundary which is set by the required pre-departure energy of the vehicles and the energy present in newly arriving EVs. The power boundary is the sum of the rated power of all the chargers with an EV currently connected. All chargers are assumed to be fully capable of V2G services, meaning the power boundary equally constrains power being charged from the grid as well as power being discharged back to the grid.

\begin{figure}[t]
    \centering
    \includegraphics[width=\linewidth]{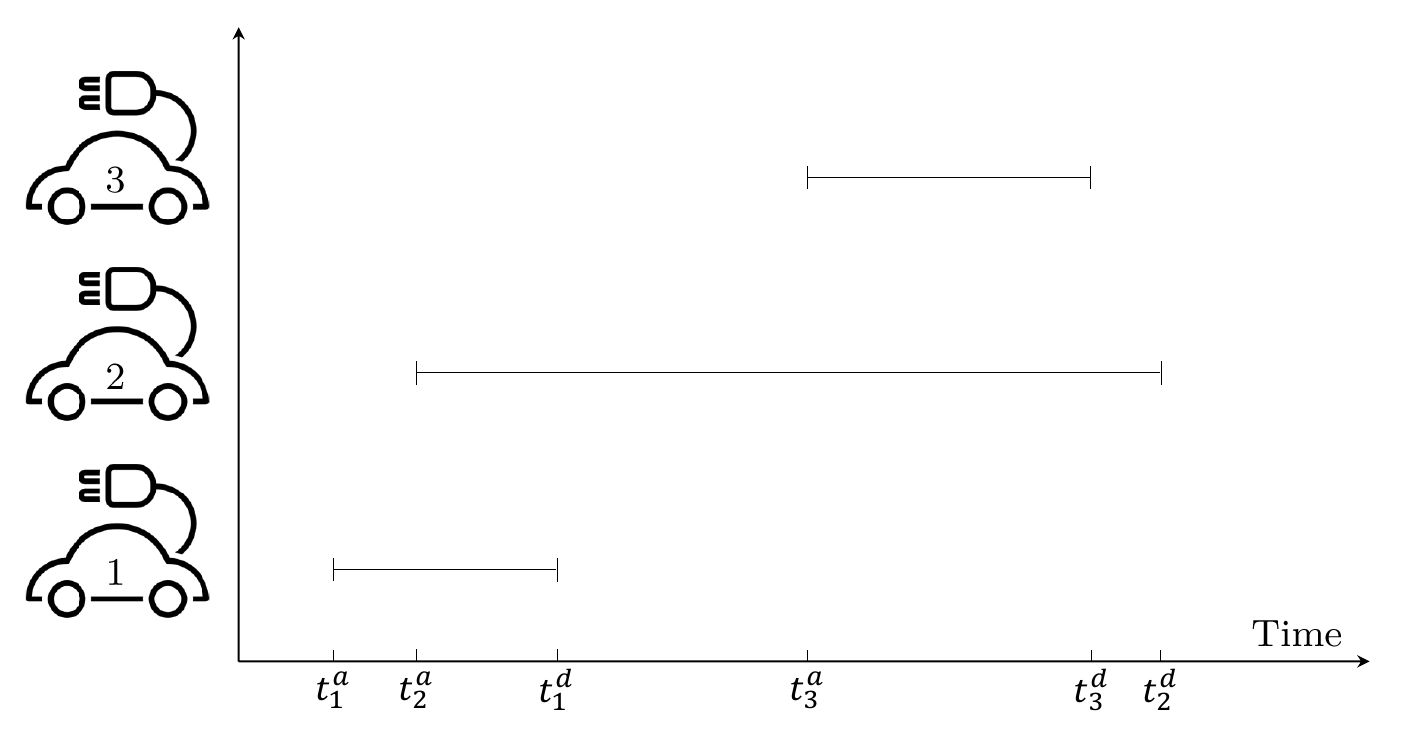}
    \caption{Arrival and Departure Times for an Example EV Fleet of 3 Vehicles}
    \label{fig:evtimes}
\end{figure}

\begin{figure}[t]
\begin{center}
\begin{scaletikzpicturetowidth}{(\linewidth - 2em)/2}
\begin{tikzpicture}[scale=\tikzscale]
    \begin{axis}[xlabel = \Large Time, ylabel = \Large Power (kW), axis y line = left, axis x line = center, ticks = none, xmin=0, xmax=65, ymin=-15, ymax=15, clip=false]
    \addplot [color = ForestGreen, very thick] coordinates {(5,0) (5,2)} ;
    \addplot [color = ForestGreen, very thick] coordinates {(5,2) (10,2)} ;
    \addplot [color = ForestGreen, very thick] coordinates {(10,2) (10,10)} ;
    \addplot [color = ForestGreen, very thick] coordinates {(10,10) (20,10)} ;
    \addplot [color = ForestGreen, very thick] coordinates {(20,10) (20,8)} ;
    \addplot [color = ForestGreen, very thick] coordinates {(20,8) (35,8)} ;
    \addplot [color = ForestGreen, very thick] coordinates {(35,8) (35,13)} ;
    \addplot [color = ForestGreen, very thick] coordinates {(35,13) (55,13)} ;
    \addplot [color = ForestGreen, very thick] coordinates {(55,13) (55,8)} ;
    \addplot [color = ForestGreen, very thick] coordinates {(55,8) (60,8)} ;
    \addplot [color = ForestGreen, very thick] coordinates {(60,8) (60,0)} ;
    \addplot [color = red, very thick] coordinates {(5,0) (5,-2)} ;
    \addplot [color = red, very thick] coordinates {(5,-2) (10,-2)} ;
    \addplot [color = red, very thick] coordinates {(10,-2) (10,-10)} ;
    \addplot [color = red, very thick] coordinates {(10,-10) (20,-10)} ;
    \addplot [color = red, very thick] coordinates {(20,-10) (20,-8)} ;
    \addplot [color = red, very thick] coordinates {(20,-8) (35,-8)} ;
    \addplot [color = red, very thick] coordinates {(35,-8) (35,-13)} ;
    \addplot [color = red, very thick] coordinates {(35,-13) (55,-13)} ;
    \addplot [color = red, very thick] coordinates {(55,-13) (55,-8)} ;
    \addplot [color = red, very thick] coordinates {(55,-8) (60,-8)} ;
    \addplot [color = red, very thick] coordinates {(60,-8) (60,-0)} ;
    \addplot[very thick, black]coordinates{(5,0)(5,0.5)};
    \addplot[very thick, black]coordinates{(10,0)(10,0.5)};
    \addplot[very thick, black]coordinates{(20,0)(20,0.5)};
    \addplot[very thick, black]coordinates{(35,0)(35,0.5)};
    \addplot[very thick, black]coordinates{(55,0)(55,0.5)};
    \addplot[very thick, black]coordinates{(60,0)(60,0.5)};
    \node [color=black] [below] at (5,0){\Large $t_1^a$};
    \node [color=black] [below] at (10,0){\Large $t_2^a$};
    \node [color=black] [below] at (20,0){\Large $t_1^d$};
    \node [color=black] [below] at (35,0){\Large $t_3^a$};
    \node [color=black] [below] at (55,0){\Large $t_3^d$};
    \node [color=black] [below] at (60,0){\Large $t_2^d$};
    \addplot[very thick, black, dashed]coordinates{(5,0)(5,1)(12,1)(12,-7)(15,-7)(15,0)(22,0)(22,4)(35,4)(35,0)(55,0)(55,4.4)(60,4.4)(60,0)};
    \end{axis}
\end{tikzpicture}
\end{scaletikzpicturetowidth}
\qquad
\begin{scaletikzpicturetowidth}{(\linewidth - 2em)/2}
\begin{tikzpicture}[scale=\tikzscale]
    \begin{axis}[xlabel = \Large Time, ylabel = \Large Energy (kWh), axis y line = left, axis x line = center, ticks = none, xmin=0, xmax=70, ymin = -50, ymax = 70,axis x line shift =0, clip=false]
    \addplot [very thick, ForestGreen] coordinates {(5,0)(10,10)(15,50)(35,50)(37,60)(60,60)};
    \addplot[very thick, red] coordinates{(5,0)(8,-6)(10,-6)(12,-32)(15,-40)(20,-30)(35,-30)(38,-45)(50,-45)(55,20)(60,60)};
    \addplot[very thick, black]coordinates{(5,0)(5,2)};
    \addplot[very thick, black]coordinates{(10,0)(10,2)};
    \addplot[very thick, black]coordinates{(20,0)(20,2)};
    \addplot[very thick, black]coordinates{(35,0)(35,2)};
    \addplot[very thick, black]coordinates{(55,0)(55,2)};
    \addplot[very thick, black]coordinates{(60,0)(60,2)};
    \node [color=black] [below] at (5,0){\Large $t_1^a$};
    \node [color=black] [below] at (10,0){\Large $t_2^a$};
    \node [color=black] [below] at (20,0){\Large $t_1^d$};
    \node [color=black] [below] at (35,0){\Large $t_3^a$};
    \node [color=black] [below] at (55,0){\Large $t_3^d$};
    \node [color=black] [below] at (60,0){\Large $t_2^d$};
    \addplot[very thick, black, dashed]coordinates{(5,0)(12,7)(15,-14)(22,-14)(35,38)(55,38)(60,60)};
    \end{axis}
\end{tikzpicture}
\end{scaletikzpicturetowidth}
\end{center}
\caption{Power and Energy Boundaries for Example EV Fleet from Figure~\ref{fig:evtimes}, Including a Possible Charging Trajectory (Black Dashed Line)}
\label{fig:aggbounds}
\end{figure}
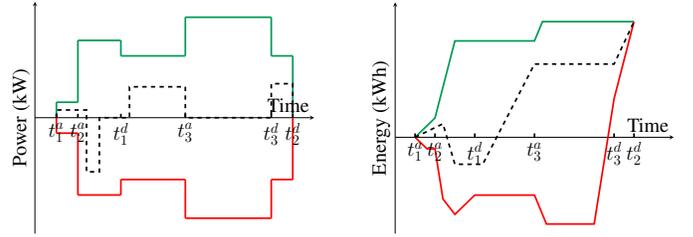

Charging can be constrained solely by the shown boundaries, so any charging trajectory that stays between the two energy boundaries and respects the power boundary should be physically possible and would ensure that departing EVs leave fully charged. Arrival and departure times for an exemplary EV fleet are shown in Figure~\ref{fig:evtimes} with the corresponding boundaries displayed in Figure~\ref{fig:aggbounds}. The aggregate boundaries ensure that any compliant charging trajectory honours the physical limitations of each battery and charger while ensuring each battery is fully charged upon departure. For further details on the aggregation procedure and the resulting boundaries, please refer to \cite{7463483} as the work in this present paper builds on their approach.

\subsection{Requirement for Energy Transfers between EVs within Aggregate Battery} 
While any charging trajectory within the boundaries is possible, \cite{7463483} does not mention that this only applies under one important assumption: the plugged-in cars are able to transfer electrical energy to and from each other via the grid in what could be described as vehicle-to-grid-to-vehicle (V2G2V) charging. The reason is that the energy boundaries only ensure that, before the departure of a vehicle, enough energy is present in the virtual battery to charge the departing vehicle. They do not dictate which vehicle the energy is in and when it is charged to the virtual battery. It may be that the virtual battery is charged ahead of a vehicle arriving and the charge has to be transferred to that vehicle from another vehicle after its arrival. This is demonstrated by the example charging trajectory in Figure~\ref{fig:aggbounds} which shows no power is delivered to the virtual battery between the arrival and departure of vehicle 3. To fulfil the charging requested by vehicle 3, vehicle 2 has previously acquired that energy and later transferred it to vehicle 3 via the grid. 

Energy transfers between EVs are a realistic possibility, even for distributed fleets, as long as chargers are physically capable of bidirectional charging but several issues need consideration such as power flow limitations, the required communications, and the revenue transactions \cite{shafiqurrahman2023vehicle}. It is important to note that an aggregate battery that exploits intra-fleet energy transfers to obtain greater flexibility, as implied by \cite{7463483}, incurs some power losses and perhaps use-of-system charges for the grid. Allowing the use of energy transfers averts the need for the additional constraints noted by \cite{wang2022integrating} that reintroduce consideration of individual EVs into the aggregate model and thereby annul its simplification. Estimating the amount of energy transferred between vehicles, required for the aggregate model from \cite{7463483} to function, is an important piece of future work because it will affect the accuracy of the energy trajectory and thereby the model's efficacy for EV scheduling.

\subsection{SOC-related Power Fluctuations} 
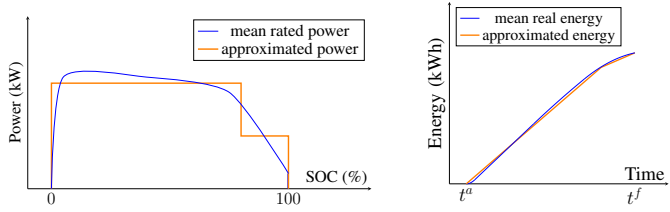
\begin{figure}[t]
\begin{center}
\begin{scaletikzpicturetowidth}{(\linewidth - 2.1em)/5*3}
\begin{tikzpicture}[scale=\tikzscale]
    \begin{axis}[xlabel = \Large SOC (\%), ylabel = \Large Power (kW), axis y line = left, axis x line = center, ticks = none, xmin=0, xmax=29, ymin=0, ymax=17, clip=false, x = 0.35cm,y = 0.3cm]
    \addplot[color = blue] coordinates {(21.95,1.5)(21.95,0)};
    \addplot [color = orange, very thick] coordinates {(2,0) (2,10.4)(18,10.4)(18,5.2)(22,5.2)(22,0)} ;
    \draw [blue] plot [smooth, tension=0.4,very thick] coordinates {(2,0) (3,11) (10,11)(17,9.5)(22,1.5)};
    \node [color=black] [below] at (2,0){\Large $0$};
    \node [color=black] [below] at (22,0){\Large $100$};
    \legend{\Large mean rated power, \Large approximated power};
    \end{axis}
\end{tikzpicture}
\end{scaletikzpicturetowidth}
\qquad
\begin{scaletikzpicturetowidth}{(\linewidth - 2.1em)/5*2}
\begin{tikzpicture}[scale=\tikzscale]
    \begin{axis}[xlabel = \LARGE Time, ylabel = \LARGE Energy (kWh), axis y line = left, axis x line = center, ticks = none, xmin=0, xmax=26, ymin=0, ymax=25, clip=false,legend pos=north west]
    \addplot [domain=2:3,blue]{((11/2)*(x^2) -22*x+22)/10} ;
    \addplot [color = orange, very thick] coordinates {(2,0) (18,16)(22,18)} ;
    \addplot [domain=3:10, blue]{(11*x-27.5)/10} ;
    \addplot [domain=10:16, blue]{(-(1/8)*(x^2)+13.5*x-40)/10} ;
    \addplot [domain=16:22, blue]{(-(7/12)*(x^2)+(28+(1/6))*x-(157+(1/3)))/10} ;
    \node [color=black] [below] at (2,0){\LARGE $t^a$};
    \node [color=black] [below] at (22,0){\LARGE $t^f$};
    \legend{\Large mean real energy, \Large approximated energy};
    \end{axis}
\end{tikzpicture}
\end{scaletikzpicturetowidth}
\end{center}
\caption{Approximation of charging power variation with SOC} \label{fig:SOCpower}
\end{figure}

The power with which an EV can be charged is often reduced as the battery approaches full charge and its cell's voltages begin to rise \cite{powerSOC}. This is illustrated in figure~\ref{fig:SOCpower} which also shows how these fluctuations in charging power capability can be approximated by simply dividing the charging process into two steps.

\begin{figure}[t]
\begin{center}
\begin{scaletikzpicturetowidth}{(\linewidth - 2.1em)/2}
\begin{tikzpicture}[scale=\tikzscale]
    \begin{axis}[xlabel = \Large Time, y label style={at={(axis description cs:-0.09,0.52)}}, ylabel = \Large Power (kW), axis y line = left, axis x line = center, ticks = none, xmin=0, xmax=65, ymin=-45, ymax=45, clip=false]
    \addplot [color = ForestGreen, very thick] coordinates {(10,25) (50,25)} ;
    \addplot [color = ForestGreen, very thick]  coordinates {(10,0) (10,25)} ;
    \addplot [color = ForestGreen, very thick]  coordinates {(50,25) (50,0)} ;
    \addplot [color = red, very thick]  coordinates {(10,-25) (42,-25)} ;
    \addplot [color = red, very thick]  coordinates {(10,0) (10,-25)} ;
    \addplot [color = red, very thick]  coordinates {(42,-25) (42,24.8)} ;
    \addplot [color = red, thin]  coordinates {(42,24.8) (49.8,24.8)} ;
    \addplot [color = red, thin]  coordinates {(49.8,24.8)(49.8,0)} ;
    \addplot [very thick,black] coordinates {(10,0)(10,2)};
    \addplot [very thick,black] coordinates {(50,0)(50,2)};
    \addplot [very thick,black] coordinates {(-1,25)(1,25)};
    \addplot [very thick,black] coordinates {(-1,-25)(1,-25)};
    \node [color=black] [below right] at (10,0){\Large $t^a$};
    \node [color=black] [below right] at (50,0){$t^d$};
    \node [color=black] [left] at (0,25){\Large $p^{max}$};
    \node [color=black] [left] at (0,-25){\Large $-p^{max}$};
    \node [color=black] [left] at (0,0){\Large $0$ };
    \end{axis}
\end{tikzpicture}
\end{scaletikzpicturetowidth}
\qquad
\begin{scaletikzpicturetowidth}{(\linewidth - 2.1em)/2}
\begin{tikzpicture}[scale=\tikzscale]
    \begin{axis}[xlabel = \Large Time, y label style={at={(axis description cs:-0.1,0.4)}}, ylabel = \Large Energy (kWh), axis y line = left, axis x line = center, ticks = none, xmin=0, xmax=65, ymin = -20, ymax = 50,axis x line shift =20, clip=false]
    \addplot [very thick,black] coordinates {(5,-19)(5,-21)};
    \addplot [very thick,black] coordinates {(52.5,-19)(52.5,-21)};
    \addplot[very thick, black, dashed] coordinates {(-1,-15)(12.5,-15)};
    \addplot[very thick, black, dashed] coordinates {(-1,30)(20,30)};
    \addplot[very thick, black, dashed] coordinates {(-1,0)(5,0)};
    \addplot[very thick, red] coordinates{(5,0) (12.5,-15)};
    \addplot[very thick, red] coordinates{(12.5,-15) (30,-15)};
    \addplot[very thick, red] coordinates{(30,-15) (52.5,30)};
    \addplot[very thick, red] coordinates{(52.5,29.8) (62.5,39.8)};
    \addplot[very thick, ForestGreen] coordinates{(5,0) (20,30)};
    \addplot[very thick, ForestGreen] coordinates{(20,30) (52.5,30)};
    \addplot[very thick, ForestGreen] coordinates{(52.5,30.2) (62.5,40.2)};
    \node [color=black] [below] at (5,-20){\Large $t^a$};
    \node [color=black] [below] at (52.5,-20){\Large $t^d$};
    \node [color=black] [left] at (0,0){\Large $e^a$};
    \node [color=black] [left] at (0,-15){\Large $e^{min}$};
    \node [color=black] [left] at (0,30){\Large $e^{max}$};
    \end{axis}
\end{tikzpicture}
\end{scaletikzpicturetowidth}
\end{center}
\caption{Individual EV Power and Energy Boundaries with Direct Charging at the End}
\label{fig:powenconstfluc}
\end{figure}
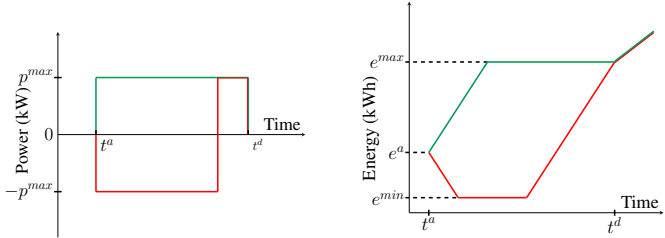

For the last 20\% of SOC, the approximation is that the battery can charge with only half of the average charging power. It would not make sense to use these final 20\% portions for flexible charging as it would take much longer to charge them than to discharge them. Simultaneously, long-lasting high SOCs are found to be harmful to battery health \cite{schmalstieg2014holistic,collath2022aging,edge2021lithium} and it, therefore, makes sense to leave the slower-charging last 20\% until the end of the charging process anyway. Albeit limited, this presents a methodology for including battery health measures in aggregate V2G models, without having to revert to the individual EV level and thereby reversing the aggregation benefits. The approximation of available power from Figure~\ref{fig:SOCpower} can be accommodated in the aggregate boundary model by simply finishing the flexible charging window earlier and assuming direct charging for the final 20\%. The resulting individual boundaries for this are shown in Figure~\ref{fig:powenconstfluc}. Note, that the direct charging interval, which is denoted by the two boundaries taking the same value, is not included in the aggregation procedure. Naturally, the direct charging intervals cannot be used for flexible charging.

\section{Stochastic Model Predictive Control} \label{smpcmodel}
The aggregate boundaries permit the assessment of the potential for flexible charging from an EV fleet but in the resesarch reported in the literature they have been deterministic, meaning they do not account for uncertainty in driving behaviour. When scheduling flexible charging for future ancillary service commitments, the availability and SOC of each vehicle will be unknown. Since the proposed boundaries are an aggregate measure and thereby converge to average behaviour, their historic evolution can be used to make predictions about their future. To account for imprecise predictions of the aggregate charging behaviour, prediction errors observed in the past are used to create a range of scenarios rather than a single prediction. Each scenario represents a possible path for the boundaries to take based on aggregate charging. The scenarios form the basis for the stochastic optimisation.

Classical model predictive control (MPC) is a control algorithm that, at each time step, optimises a process over a finite time horizon, using information from a dynamic process model \cite{rawlings2017model,mesbah2016stochastic}. Stochastic MPC (SMPC) is a robust formulation of this approach which ensures that the optimisation probabilistically considers a range of potential scenarios  \cite{king2012modeling}. This makes SMPC very well suited for real-time scheduling paired with day-ahead service auctions. Its stochastic aspect allows it to incorporate the uncertain nature of day-ahead ancillary service provision into the control actions. Furthermore, model-predictive control is well suited for real-time charge scheduling because it allows for regular decisions to be made that take the constantly updating forecasts into account. The procurement and execution of ancillary services (such as power reserve) in most power systems happen on two different timescales, so a two-stage SMPC is proposed here. The first stage considers the time interval in which the reserve service is procured and the second stage considers the intervals in which service commitments have been given but the charging decisions remain to be made. Procuring the reserve services carries an element of risk evaluation in that if the EV aggregator overcommits to providing a service it may incur a large penalty from the system operator for failure to deliver. The risk measure Conditional Value-at-Risk (CVaR) is introduced to control the risk aversion of the algorithm. CVaR allows the aggregator to optimise the expected outcome of the worst-case scenarios which are likely those in which a penalty would be incurred.

\subsection{Predictive Model and Scenario Generation}
A multiple linear regression (MLR) model is proposed to forecast each boundary. Time-of-day was identified as the most influential predictor so a separate regression was carried out for each settlement period in the day. By separately creating predictions for each settlement period in a day, the prediction model is thus informed by the historical data in that settlement period, i.e. previous arrival/departure trends at that time of day. This allows it to include the effect of environmental variables, such as precipitation ($W_t$) and temperature ($H_t$), which affect driving patterns, and thereby the boundary trajectories. 
\begin{equation}
    \label{eq1}
    E^{u}_t = \beta_{0,t} + \beta_{1,t}E^d_{t=0} + \beta_{2,t} H_t + \beta_{3,t} W_t + \beta_{4,t} B_t + \beta_{5,t} D_t + \varepsilon_t
\end{equation}

Equation (\ref{eq1}) shows the regression equation for the upper energy boundary and the same equation was applied to predict the power boundary. If the lower energy boundary had been forecast independently in the same way, there would likely be scenarios in which the lower boundary would have been predicted to be higher than the upper boundary. This would have led to an impossible optimisation task in that scenario because the charging trajectory always needs to stay below the upper boundary and above the lower boundary. To prevent an infeasible optimisation problem, the difference between the lower and upper energy boundary was predicted, instead of predicting the lower boundary directly. Since there should be no historical data of the difference between the upper and lower boundary being negative, all scenarios for the boundary difference ($E^d_t$) can be assumed positive, meaning that estimating the lower boundary through Equation~\ref{eq2} should ensure that the lower boundary is never forecast to be higher than the upper boundary. Note, that this approach does not consider any potential correlations between the boundaries. Including correlations may further enhance predictive performance and is a subject for future work.

\begin{equation}
    \label{eq2}
    E^{l}_t = E^u_t - E^d_t
\end{equation}

\begin{figure}
    \centering
    \includegraphics[width=\columnwidth]{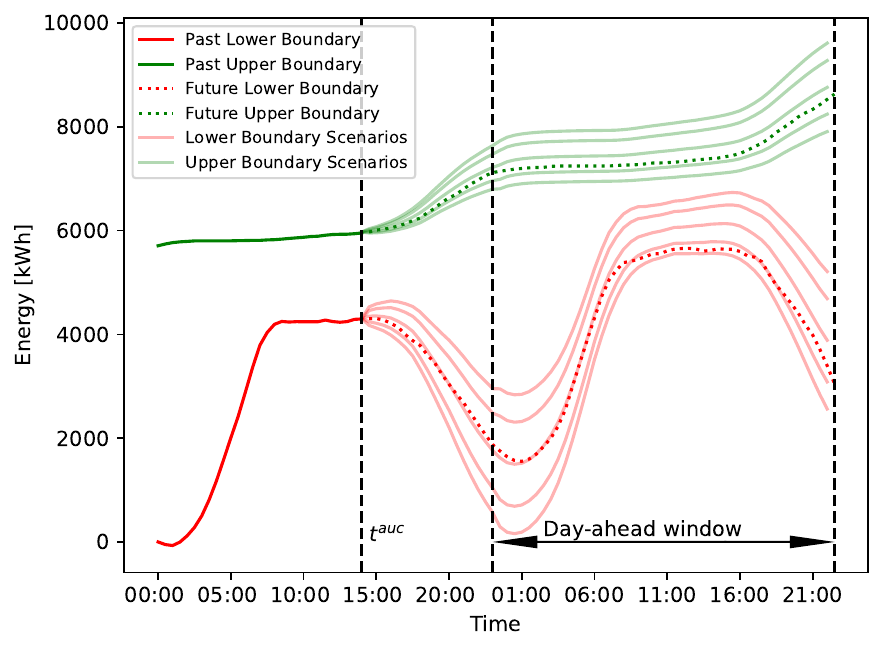}
    \caption{Predicted Scenarios for Energy Boundaries}
    \label{fig:predenbounds}
\end{figure}

\begin{figure}
    \centering
    \includegraphics[width=\columnwidth]{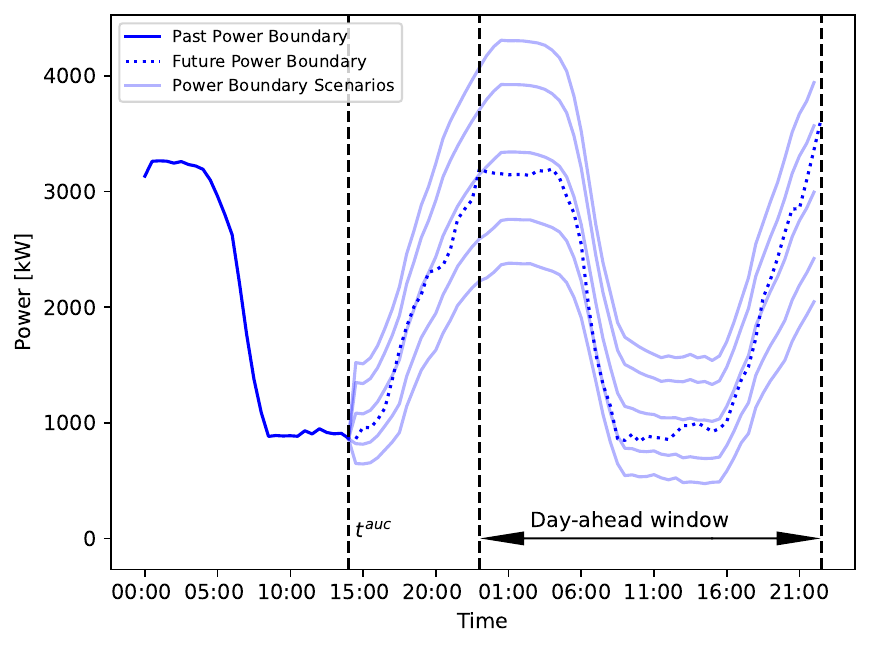}
    \caption{Predicted Scenarios for Power Boundaries}
    \label{fig:predpobounds}
\end{figure}

Scenarios are generated by fitting a normal distribution to the residuals between predictions and training data. The resulting distribution is then split into sections with different probabilities ($p_s$), i.e. $p_s$ is the frequency with which the prediction model over- or underestimates by the amount that corresponds to that scenario. To properly include the extreme ends of the distribution while using only a few scenarios for computational efficiency, different values were chosen for $p_s$ (1\%, 10\%, 78\%, 10\%, 1\%). Figures~\ref{fig:predenbounds} and~\ref{fig:predpobounds} depict some exemplary scenario predictions where each of the opaque lines represents a scenario into which the actual future boundary may fall with probability $p_s$. The scenarios are generated at the time of the reserve auction (14:00) and span the time until the end of the delivery window (23:00 - 23:00) on the next day.

\subsection{Stochastic Optimisation}

\subsubsection{Reserve bidding (First-Stage)}
To effectively contract the right amount of reserve service, the algorithm has to consider the entire time window over which the services will be delivered ($t^{auc} - T$). For this time window, the proposed algorithm weighs up the revenue from reserve services ($R_{res}$) against the risk of incurring a penalty for non-delivery ($C_{pen}$). In addition, the algorithm considers the optimal charging trajectory and the resulting cost of charging ($C_{cha}$). $C_{cha}$ and $C_{pen}$ differ for each scenario as they are affected by the future charging schedule. $R_{res}$ is determined ahead of time and is therefore the same in all scenarios. These considerations can be formulated as a stochastic mixed-integer linear programming (MILP) optimisation:

\begin{equation}
    \maximize_{ \substack{P^{PR}_{t}, P^{NR}_t, P^{G2V}_{s,t}, P^{V2G}_{s,t} \\ \Delta P^{pen}_{s,t}, b^{PR}_t, b_t^{NR}}} \quad  R^{res} - \sum_{s \epsilon S} p_s (C^{cha}_s + C^{pen}_s -A_s)\label{eq3}
\end{equation} 

Reserve revenue ($R^{res}$), charging costs ($C^{cha}_s$) and penalty payments ($ C^{pen}_s$) are defined in Equations (\ref{eq4}) - (\ref{eq6}) and Equation (\ref{eq6a}) defines an "end-of-plan credit" term ($A_s$) that compensates the algorithm for any energy that is left in the vehicles' batteries at the end of the planning window. This term alleviates the need for complex cross-window relaxations as introduced by the authors of \cite{10415236}. Equation (\ref{eq7}) defines the aggregate charging trajectory, which is referenced by Equation (\ref{eq6a}), as a function of power charged to and from the grid ($P^{V2G,g}_{s,t}$/$P^{G2V,g}_{s,t}$). 

\begin{align}
    R^{res} = \sum^{T}_{t=0} (c_t^{PR} P^{PR,g}_t + c_t^{NR} P^{NR,g}_t) \label{eq5} \\
    C^{cha}_s = \sum^{T}_{t=0} c^E_t \left(P^{G2V,g}_{s,t} - P^{V2G,g}_{s,t}\right)\Delta t \label{eq4} \\
    C^{pen}_s = \sum^{T}_{t=0} c^{pen} \Delta P_{s,t}^{pen} \label{eq6}\\
    A_s =(E^{EV}_{s,T}-E^l_{s,T})\bar{c}^E \label{eq6a}\\
    E^{EV}_{s,t_x} =  E_{t^{auc}-1}^{EV} + \sum_{t^{auc}}^{t_x-1}\left(P^{G2V,g}_{s,t}\eta-\frac{P^{V2G,g}_{s,t}}{\eta}\right)\Delta t\label{eq7}
\end{align}

The following constraints ensure that the charging trajectory stays within the power and energy boundaries and that a penalty is incurred if the contracted reserve ($P^{PR,g}_t$/$P^{NR,g}_t$) cannot be serviced:
\begin{align}
    E^u_{s,t} \geq E^{EV}_{s,t} \geq E^l_{s,t} \label{eq8}\\
    \frac{P^{V2G,g}_{s,t}}{\eta} + P^{G2V,g}_{s,t} \leq P^b_{s,t}\label{eq9}\\
    P^{PR,v}_{s,t} = \frac{P^{PR,g}_t - \frac{1}{z} \sum^{t-1}_{t-z}(P^{G2V,g}_{s,t} - P^{V2G,g}_{s,t})}{\eta} \label{eq11}\\
    P^{NR,v}_{s,t} = \left(P^{NR,g}_t + \frac{1}{z} \sum^{t-1}_{t-z}\left(P^{G2V,g}_{s,t} - P^{V2G,g}_{s,t}\right)\right)\eta \label{eq12}\\
    b^{PR}_t \times M_1 \geq P^{PR,g}_t \label{eq13} \\
    \vspace{1mm}
    b^{NR}_t \times M_1 \geq P^{NR,g}_t \label{eq14}
\end{align}
\vspace{-3mm}
\begin{equation}
    E^{EV}_{s,t} - \left(P^{PR,v}_{s,t} + \frac{\Delta P_{s,t}^{pen}}{\eta}\right)\Delta t_r \geq 
    E^l_{s,t} - (1-b^{PR}_t)M_2 \label{eq15}
\end{equation}
\vspace{-1mm}
\begin{equation}
    E^{EV}_{s,t} + (P^{NR,v}_{s,t} \eta - \Delta P_{s,t}^{pen} \eta)\Delta t_r \leq E^u_{s,t} + (1-b^{NR}_t)M_2 \label{eq16}
\end{equation}
\vspace{-1.5mm}
\begin{equation}
    \hspace{10000pt minus 1fil} P^{PR,v}_{s,t} - \frac{\Delta P_{s,t}^{pen}}{\eta} \leq P^b_{s,t} + \left(1-b^{PR}_t\right)M_2 \hfilneg \label{eq17} \\
\end{equation}
\vspace{-2mm}
\begin{equation}
    \hspace{10000pt minus 1fil} \frac{P^{NR,v}_{s,t}}{\eta} - \Delta P_{s,t}^{pen} \leq P^b_{s,t} + \left(1-b^{PR}_t\right)M_2  \hfilneg  \label{eq18}
\end{equation}

Equations (\ref{eq8}) - (\ref{eq9}) ensure that the charging trajectory ($E^{EV}_{s,t}$) always stays within the energy and power boundaries. Since power charged from the grid, and power fed back to the grid are considered separate positive entities ($P^{V2G,g}_{s,t},P^{G2V,g}_{s,t}$), the power boundary constraint can be expressed as a single inequality with the sum of the two entities required to remain below the power boundary. Equations (\ref{eq11}) and (\ref{eq12}) express the reserve power that has to be provided by the EV fleet as a function of the contracted reserve power and the baseline demand/generation from the previous $z$ time intervals. Without further adjustment, this would mean that, in the case of no reserve commitments ($P^{PR,g}_t = 0$), the fleet would be constrained to provide reserve power based on its baseline demand ($P^{PR,v}_t = - \sum^{t-1}_{t-z}(P^{G2V,g}_{s,t} - P^{V2G,g}_{s,t})$). Naturally, there should be no reserve constraints if the aggregator has not committed to providing any reserve power. Thus, Equations (\ref{eq13}) and (\ref{eq14}) introduce a binary variable, $b_t$, that can only take the value zero if no reserve has been scheduled for that particular time interval $t$. Inevitably, introducing $b_t$ turns the optimisation into a MILP problem, increasing its complexity. Equations (\ref{eq15}) and (\ref{eq16}) ensure that a penalty is applied if the reserve commitments violate the energy boundaries and Equations (\ref{eq17}) and (\ref{eq18}) ensure the same for the power boundary. In the case of no reserve commitments ($b_t = 0$), the term $(1-b_t)M_2$ adds a sufficiently large number to the respective boundary to make sure it is not limiting. Sufficiently large for $M_1$ means it needs to be larger than any value that $P^{PR,g}_t$ or $P^{NR,g}_t$ may take. This is ensured by taking a multiple of the maximum possible difference between upper and lower energy boundary. $M_2$ ensures that boundaries do not constrain the non-zero baseline reserve ($P^{PR,v}_t$) when no reserve commitments ($P^{PR,g}_t$) have been made. Equation (\ref{eq11}) shows $P^{PR,v}_t$ is a function of only $P^{PR,v}_t$ and the baseline power from the previous $z$ intervals. $M_2$, therefore, needs to only be larger than the largest possible baseline power term to ensure that the boundaries do not constrain the charging trajectory when no reserve is scheduled.

\subsubsection{Charge Scheduling (Second-Stage)}
With the reserve provision for a future time window already procured, for each time interval within that window, the algorithm simply needs to weigh up the charging costs ($C_{cha}$) and the expected cost of incurring a penalty ($C_{pen}$).

\begin{equation}
    \minimize_{P^{G2V}_{s,t}, P^{V2G}_{s,t}, \Delta P^{pen}_{s,t}} \quad \sum_{s \epsilon S} p_s (C^{cha}_s + C^{pen}_s -A_s) \label{eq19} \\
\end{equation}

subject to: (\ref{eq4}) - (\ref{eq12}), (\ref{eq15}) - (\ref{eq18}) \\

With no reserve service decisions to be made, the charge scheduling step only considers scenarios to find the optimal charging trajectory in all situations. Since grid-side reserve power ($P^{PR,g}_t$) is already decided in the first stage, the computationally expensive MILP variables from equations (\ref{eq13}) and (\ref{eq14}) can be dropped. Instead, a simple if-statement decides whether equations (\ref{eq15}) - (\ref{eq18}) apply for each time interval, depending on whether any grid-side reserve was scheduled in the first stage for this time interval. Since the decisions in which risk aversion plays a role were already made in the first stage, the second-stage optimisation can have the expected value as its objective. This means that the risk measure CVaR is only applied to the first-stage optimisation.

\subsubsection{Conditional Value-at-Risk}
The first-stage optimisation involves the commitment to provide a future service with only limited certainty that the resources to provide the service will actually be available. With the possibility of incurring prohibitively large penalties for non-delivery, this requires weighing up the risk of incurring a penalty against higher expected returns. Conventional stochastic programming optimises the expected value and thus fails to distinguish risk preferences. Conditional Value-at-Risk is a common risk measure that expresses the expected value in the $\alpha\%$ worst-case outcomes, i.e. $CVaR(\alpha = 0.1)$ expresses the expected value for the worst 10\% of possible outcomes. \cite{rockafellar2000optimization} introduced a methodology for including the risk measure CVaR in the objective of a stochastic programming problem while keeping the problem convex. This allows a stochastic algorithm to fine-tune its risk aversion. For our case, this is formulated following previous implementations of CVaR in a stochastic programming objective function \cite{dimanchev2023consequences,munoz2017does}:

\begin{equation}
    \min \quad (1-\Omega)\times(C^{cha}_s + C^{pen}_s - R^{res}) + \Omega \times \CVaR \label{eq20}
\end{equation}

subject to:
\vspace{-2mm}
\begin{align}
    \CVaR \geq \VaR + \frac{1}{1+\alpha} \sum_{s \epsilon S} p_s w_s \label{eq21}\\
    w_s \geq (C^{cha}_s + C^{pen}_s) - R^{res} - \VaR \label{eq22}
\end{align}

and (\ref{eq7}) - (\ref{eq18}) \\

Equation (\ref{eq18}) is the updated objective function that considers both the expected outcome and the risk measure CVaR. $\Omega$ decides the risk-aversion of the algorithm with a higher $\Omega$ increasing the weighting of $\CVaR$. Equations (\ref{eq19}) - (\ref{eq20}) constrain the $\CVaR$ following \cite{rockafellar2000optimization}. Additionally, the operational constraints from Equations (\ref{eq7}) - (\ref{eq16}) still apply. Three risk-aversion factors ($\Omega$) are considered: a risk-neutral approach ($\Omega=0$), a risk-averse approach ($\Omega=0.5$), and a least-regrets approach ($\Omega=1$) that solely focuses on the outcome in the worst case. The risk threshold is set to 10\% ($\alpha = 0.1$).

\section{Case Study} \label{casestudy}
The proposed charging algorithm was trialled in a case study on the "Quick Reserve" product by the British system operator National Grid Electricity System Operator (NG ESO). Data and code files for the case study have been made available \footnote{\url{https://github.com/ImperialCollegeLondon/EV_reserve}}, and a Gurobi solver \cite{gurobi} was used via its Python API to solve the MILP optimisation.

\subsection{NG ESO Quick Reserve Product}
Quick Reserve is an ancillary service product that NG ESO planned to introduce in October 2023 although its implementation has been delayed \cite{NGreserve}. Quick reserve will require activation periods of below 1 minute which makes it a good fit for V2G-capable EVs. The activation periods will be no longer than 15 minutes and units will be allowed a recovery period of up to 3 minutes after being activated before the units must be available again. The maximum time for which Quick Reserve could be active ($\Delta t_r$) in any half-an-hour settlement period is therefore 27 minutes (two activations separated by 3 minutes). Reserve activation is not specifically considered. The algorithm solely ensures that enough power and energy capacity is present in each half-hour settlement to service at least 27 minutes of reserve. The baseline, against which the delivered reserve will be measured, is taken from the hour (two settlement periods) prior to activation. Reserve capacities will be auctioned every day at 2 pm for the following day, divided into 2h service windows (SW) from 11 pm until 11 pm \cite{NGreserve}. Reserve commitments will have to be equal across four adjacent settlement periods. Quick reserve comprises both positive quick reserve ($P^{PR}$) providing power to the grid and negative quick reserve ($P^{NR}$) which is additional load to balance sudden load rejection elsewhere. 

\begin{figure*}
    \centering
    \includegraphics[width=\textwidth]{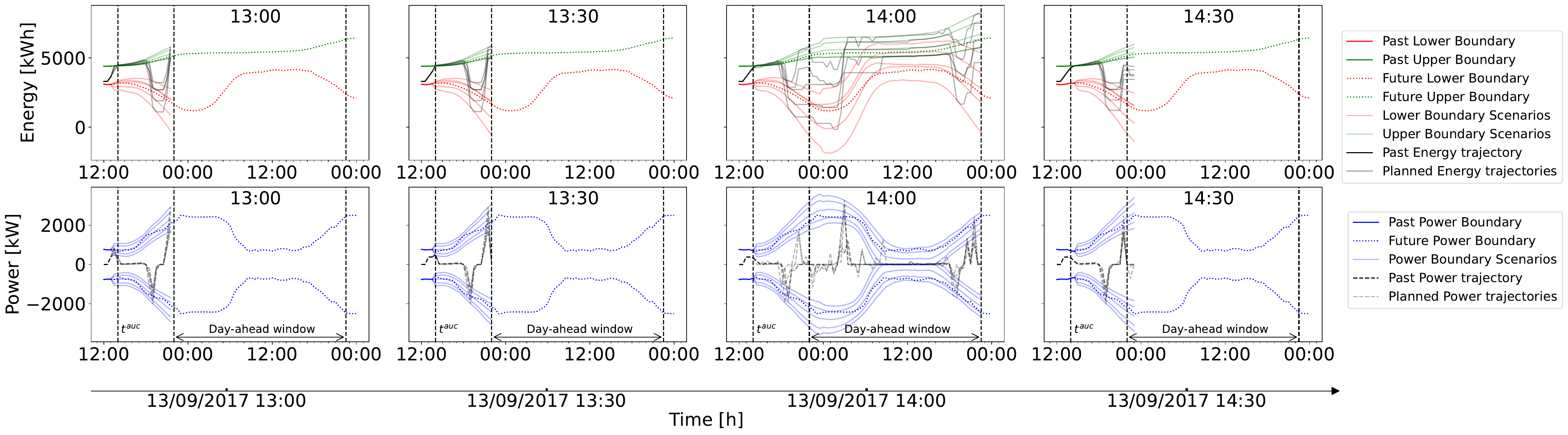}
    \caption{Illustration of SMPC Algorithm for Aggregate Energy and Power Boundaries of 720 EVs. Plans are for settlements starting 13:00 till 14:30 with day-ahead auction at 14:00 for which the auction time and delivery window are indicated in all graphs.}
    \label{fig:illuSMPC}
\end{figure*}

\subsection{Data}
\subsubsection{EV charging}
The UK Department for Transport \cite{dftEVdata} provides a large EV charging dataset comprising 25,000 chargers and 3.2 million charging events. For each charging event, the data includes the charger ID, plug-in time, plug-out time, and the amount of energy that was charged. The dataset does not contain information about the EV that was charging but, because only domestic chargers are included, it is assumed here that each charger serves only one vehicle and thus information from that vehicle can be inferred from the charger data. The rated power of each charger was assumed to be the maximum power that is seen to be drawn by that charger across the data set

\begin{equation}
    p^{max}_i = \max\left(\frac{\Delta e_j}{\Delta t_j} | j \epsilon J\right),
\end{equation}

where $J$ is the set of charging events that are completed at charger $i$. Similarly, the maximum battery capacity of each vehicle is taken as the maximum energy that is delivered to it in any charging event
\begin{equation}
    e^{max}_i = \max \left(\Delta e_j | j \epsilon J \right).
\end{equation}

Because some chargers may not have charged at their rated power, or vehicles may never have fully charged their battery in a single charge across the dataset minimum values of 7 kW and 16 kWh were set. It was also assumed that all chargers are bidirectional, which is unlikely to be the case in the actual data but was important for exploring the V2G concept. Cleansing of the raw data included the exclusion of any two overlapping charging events for the same charger ID and the exclusion of charging events that lasted for more than a week. The charging efficiency was assumed to be 90\% ($\eta = 0.9$).

\subsubsection{Wholesale electricity and reserve service prices}
Price data for the British wholesale day-ahead market is no longer published, so the system price in the Balancing Mechanism Reporting System (BMRS) was taken as a substitute \cite{elecpricedata}. Since the Quick Reserve product has not been launched yet, price data for reserve services was adapted from the old "Fast Reserve" product \cite{NG_FFR_tender}. This derived Quick Reserve price data was compared with auction data from the Synchronous Reserve product in the Pennsylvania-New Jersey-Maryland (PJM) Interconnection \cite{pjm_data} for validation. The results were a nighttime (23:00 - 07:00) positive reserve reward of \pounds0.31/MW per settlement and a daytime (07:00 - 23:00) reward of \pounds1.41/MW per settlement. Due to a lack of publicly available data for the British reserve market, negative reserve prices were assumed to statically be at 30\% of positive reserve prices. National Grid ESO states that non-delivery of ancillary services can be penalised by up to 30\% of the monthly revenue \cite{NGreserve_perf}. This was iteratively estimated to be £52/MW.

\subsubsection{Regressors in the predictive model}
Two predictive variables required additional data, precipitation ($W_t$) and temperature ($H_t$), which was sourced from the respective Hadley centre datasets \cite{alexander2000updated,parker1992new}. Note that precipitation data is for England and Wales while temperature data is for central England.

\section{Results} \label{results}

\subsection{Illustration of Algorithm Performance} \label{algoperf}
Figure~\ref{fig:illuSMPC} illustrates how the SMPC algorithm predicts scenarios and solves the resulting stochastic optimisation for every settlement period. At 13:00 and 13:30, the algorithm considers the next 9 hours (until 22:00 and 22:30 respectively) to optimise charging for that period. The algorithm considers a much longer time window at 14:00 than at other times because at this point the algorithm has to choose reserve commitments for a 24-hour period beginning at 23:00 and thus has to consider scenarios for the 33 hours ahead. With the reserve commitments for the next day decided the algorithm can then return to a 9-hour time window in the subsequent settlement period (14:30).

\begin{figure}
    \centering
    \includegraphics[width=\columnwidth]{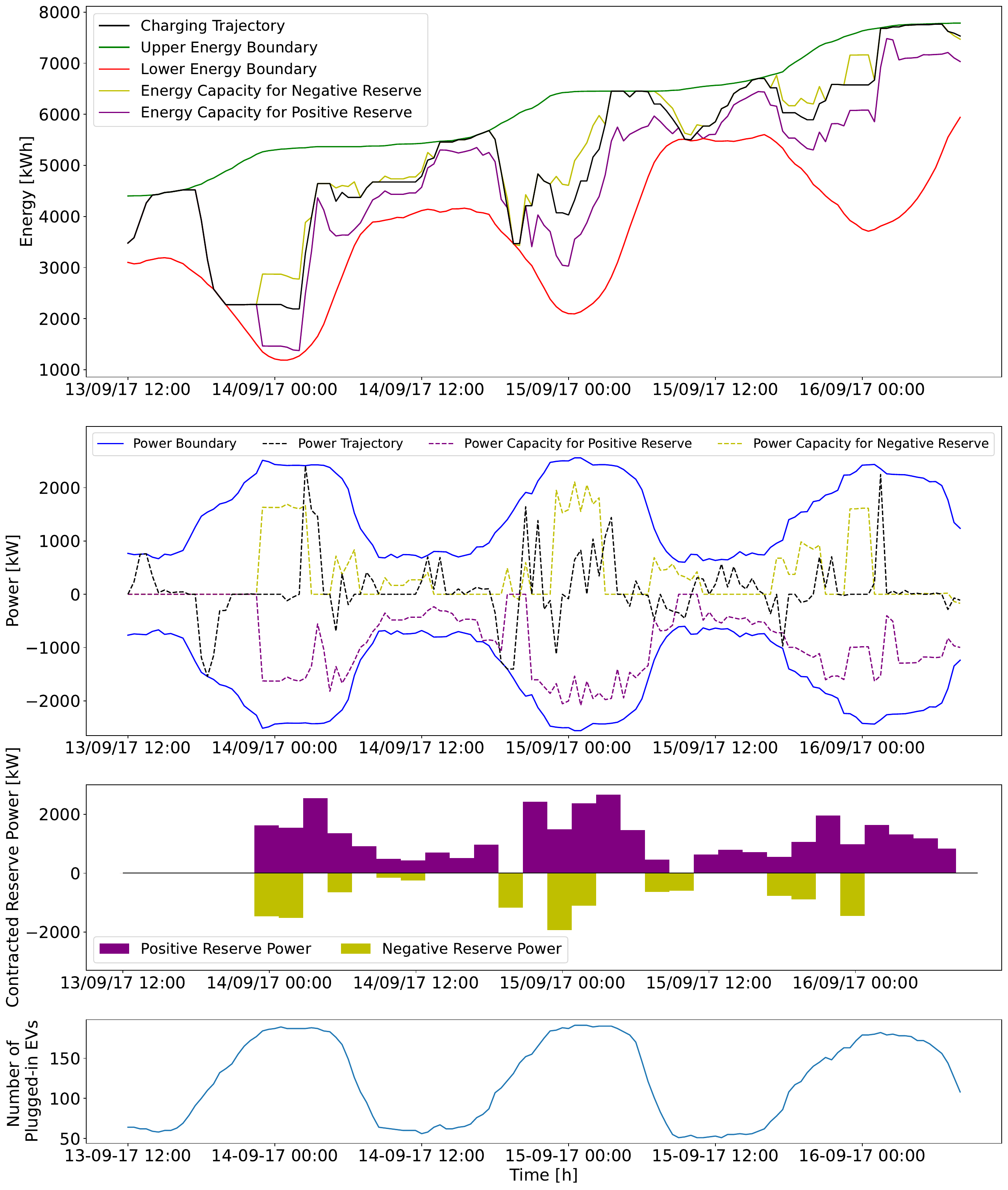}
    \caption{Illustration of Charging Trajectory and Reserve Commitments With Energy and Power Boundaries for risk-averse aggregator ($\Omega = 0.5$) of a 720 EV fleet}
    \label{fig:illu_algo}
\end{figure}

Figure~\ref{fig:illu_algo} shows the resulting trajectories that the process shown in Figure~\ref{fig:illuSMPC} produces. In yellow and purple, the capacity that has to be kept to service negative and positive reserve has been added to the power and energy trajectories. While these depend on the contracted reserve, shown in the third part of Figure~\ref{fig:illu_algo}, they do not simply equate because delivered reserve ($P^{PR,v}$) depends both on contracted reserve ($P^{PR,g}$) and the previous baseline as stated in Equations (\ref{eq11}) - (\ref{eq12}). For a given settlement, the delivered reserve is the difference between the contracted reserve and the power delivered in the two previous settlements, i.e. the baseline. Positive reserve services denote the feeding of power into the grid, reducing the energy levels in the aggregate battery. The energy capacity for positive reserve therefore falls below the charging trajectory because the aggregator has to keep sufficient energy levels to discharge positive reserve energy without reaching the lower energy boundary. Similarly, negative reserve means absorbing power from the grid, for which the aggregator has to ensure that there is sufficient "empty" battery capacity. At times, when the aggregator schedules no reserve despite sufficient capacity, this is usually because the algorithm expects greater earnings from energy arbitrage than from reserve service provision. Also note, how the risk-averse nature of the algorithm makes it systematically schedule reserve services well below the power boundaries to ensure that the likelihood of not having enough power available stays sufficiently small.

Figure~\ref{fig:illu_algo} also shows that the energy boundaries rarely constrain the reserve services and it is usually the power boundaries that dictate how much reserve power the algorithm can schedule. This observation is consistent throughout all results and indicates that the limiting factor for a vehicle to deliver reserve services is the charger power rather than the battery capacity. Charger infrastructure could therefore be seen as more important than battery energy management for the delivery of ancillary services via V2G.


\begin{figure}
    \centering
    \includegraphics[width=\columnwidth]{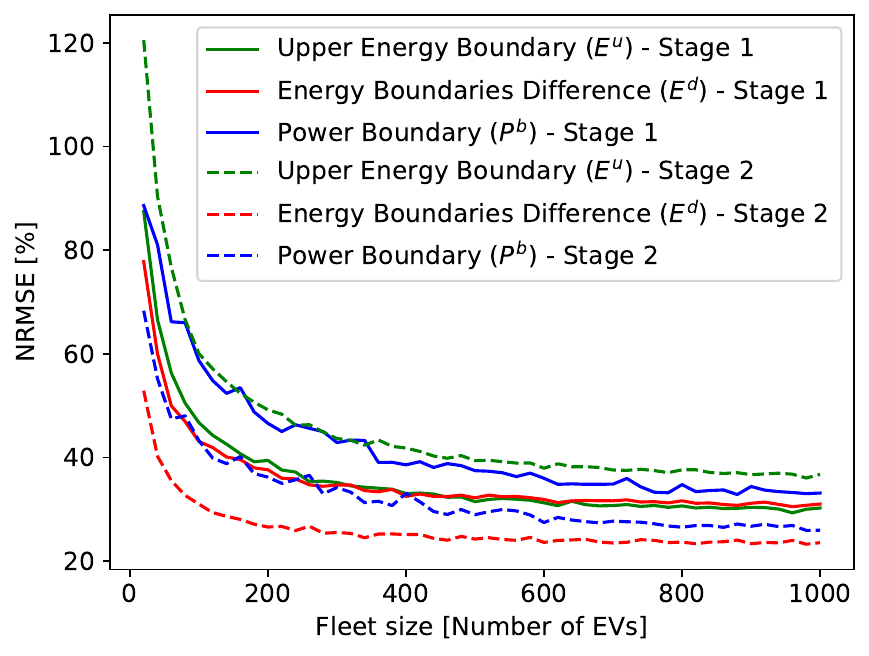}
    \caption{Coefficient of Determination ($R^2$) against Different Fleet Sizes}
    \label{fig:r2}
\end{figure}

\subsection{Prediction Performance} 
Figure~\ref{fig:r2} plots the normalised root mean squared error (NRMSE), which is calculated as

\begin{equation}
    NRMSE = \frac{\sqrt{\frac{1}{T}\sum^T_{t=1}(\hat{y}(t)-y(t))^2}}{\frac{1}{T}\sum^T_{t=1} y(t)},
\end{equation}

denoting the root mean squared error as a percentage of the average of the predicted variable. This makes the NRMSE a decent measure for comparing the performance of different prediction algorithms. Figure~\ref{fig:r2} illustrates that the NRMSE decreases with fleet size, which is as expected, as more EVs mean that the aggregate boundaries diverge less from their expected values which the MLR model attempts to predict. It can also be seen that the NRMSE for the power boundary difference and the energy boundaries difference is lower for the second stage compared to the first stage. This can be explained by the longer prediction window for the first stage compared to the second stage (33 hours vs 9 hours) which, in turn, results in lower errors for the second stage predictions. The reason that the upper boundary predictions do not follow this trend is likely to do with variations in the denominator, i.e. average value of the boundary itself. This demonstrates how NRMSE can be affected by the absolute value of the predicted variable, presenting a major limitation when comparing prediction performance. Using an autoregressive model, \cite{pertl2018equivalent} achieve a slightly lower NRMSE range of 10\% - 24\% for the day-ahead prediction of the parameters in their aggregate model. This difference may, however, be influenced by the fact that they are predicting the parameters of an entirely different aggregation model. Combining an autoregressive prediction model, such as the one from \cite{pertl2018equivalent}, with additional regressors from an MLR model, such as the ones in this present paper, promises to yield improved prediction results. In contrast to \cite{pertl2018equivalent}, the predictions in this present paper are translated into scenarios for stochastic programming, which allows for errors in prediction models to be accounted for.

\subsection{Comparison of SMPC against Benchmarks} \label{fleetsize}

\begin{figure}
    \centering
    \includegraphics[width=\columnwidth]{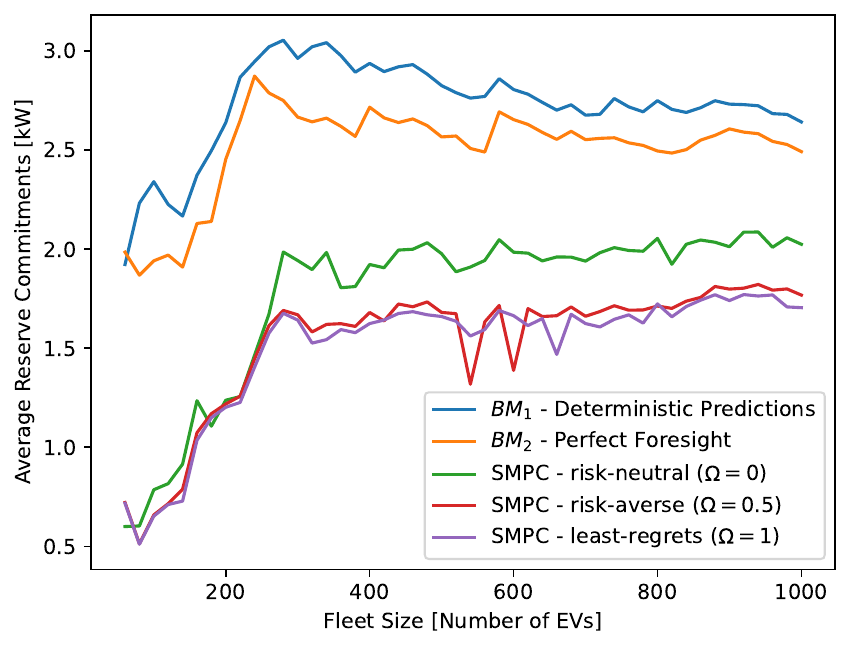}
    \caption{Average Contracted Reserve per Vehicle for Different Fleet Sizes}
    \label{fig:perev_reserve}
\end{figure}

Figure~\ref{fig:perev_reserve} plots the average reserve that can be scheduled per vehicle against fleet size for 5 algorithms. Two benchmark algorithms are shown: one that, unrealistically, has perfect future knowledge ($BM_2$) but serves as a limit case and the other ($BM_1$) is a deterministic version of our model that makes a single prediction but does not account for uncertainty and makes naive predictions of reserve that risk non-delivery penalties. Three cases for the proposed SMPC algorithm are shown for different risk preferences and all three scheduled less reserve than the benchmarks because they cannot match perfect foresight and are not naive about penalties. The risk-neutral algorithm ($\Omega=0$) optimistically schedules more reserve because it is more prone to ignore the worst scenarios. With improving prediction accuracy, as shown in Figure~\ref{fig:r2}, more reserve can be scheduled per vehicle but at fleet sizes beyond 250 EVs, the reserve commitment levels off which is likely because the predictions improve at a slower rate beyond that. Once a certain prediction quality has been achieved, further small prediction improvements are likely to have little impact on the algorithm's reserve commitments as they present only small changes to the predicted scenarios. 

\begin{figure}
    \centering
    \includegraphics[width=\columnwidth]{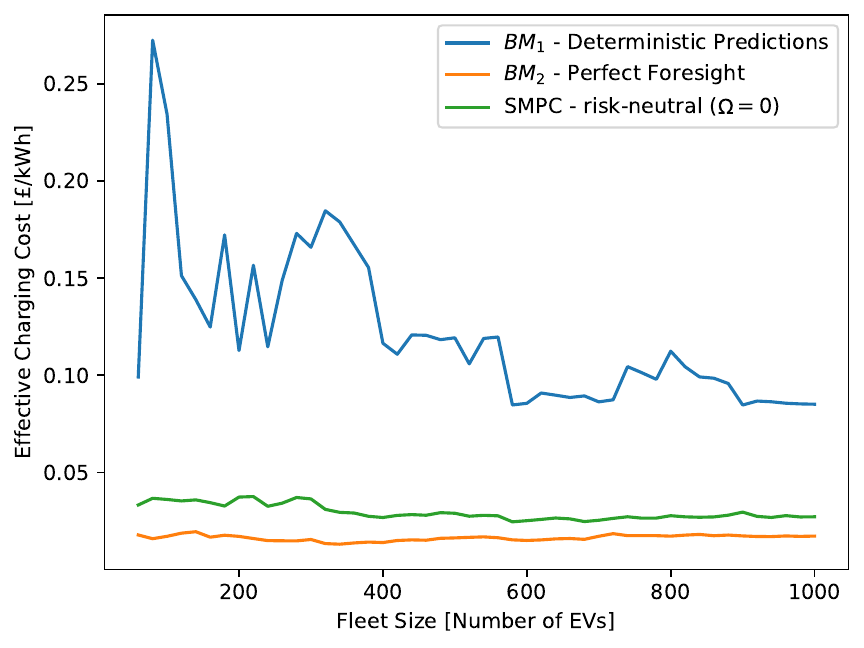}
    \caption{Effective Charging Costs ($C^{cha} + C^{pen} − R^{res}$) over Fleet Size}
    \label{fig:chargecosts}
\end{figure}

\begin{figure}
    \centering
    \includegraphics[width=\columnwidth]{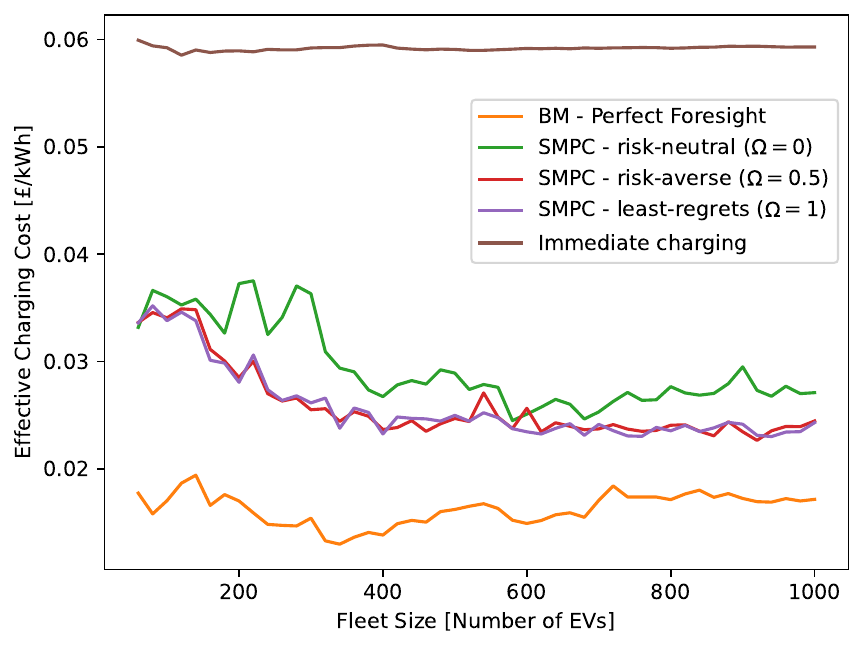}
    \caption{Effective Charging Costs For Different Risk Aversion Factors}
    \label{fig:chargecostsdetail}
\end{figure}

The total cost of charging is made up of the wholesale electricity costs minus the revenues earned from reserve contracts plus any penalty for non-delivery of that service. Figure~\ref{fig:chargecosts} plots the total cost for the risk-neutral algorithm and the two benchmarks. The perfect foresight case earns a high revenue from reserve and suffers no penalties so it achieves the lowest cost. The risk-neutral SMPC case has somewhat higher costs (lower reserve revenue and some penalties) but approaches the perfect foresight case. The simple deterministic method has the highest cost because of large penalties arising from over-commitment of reserve. This $BM_1$ simulates the algorithm from \cite{7463483} without the stochasticity that was added in this paper but any of Table~\ref{tab:litreview}'s studies without stochasticity would be expected to perform similarly in this context. The costs of three different risk versions for SMPC are compared in Figure~\ref{fig:chargecostsdetail} and are seen to have only small differences between them. All charge at lower costs,  though, than the simple case of charge-on-arrival with no services ($\sim6p/kWh$) and all approach the perfect foresight case ($\sim1.7 p/kWh$). SMPC performance continues improving for larger fleets until a fleet size of roughly 400 is reached where the effective cost plateaus at around 2.5p/kWh, some 60\% below charge-on-arrival. The reason the plateau is at 400 EVs not the 250 EVs at which per-EV reserve commitments level off, is that the incurred penalties reduce between 250 and 400 EVs due to more accurate predictions. For the stochastic algorithms, a 25\% reduction in effective charging costs can be attributed to energy arbitrage with the remaining 15 - 35 percentage points being reserve service revenue. Price reductions are based on current reserve price estimations (2.09 £/MW/h) which may change due to VRE penetration \cite{hirth2015balancing}, storage expansion \cite{schmidt2023monetizing}, or market power of a single aggregator\cite{schuler2001electricity}. The risk-neutral algorithm has somewhat higher costs than the other two SMPC cases which is due to higher penalty payments and somewhat unexpected because the risk-neutral algorithm focuses on the expected value. One possible explanation is that the in-sample scenario generation is being used for out-of-sample predictions. This may result in the scenarios overestimating the accuracy of the predictions and thereby leading to more "worst scenarios" materialising, disadvantaging the risk-neutral algorithm.

\section{Conclusion} \label{concl}
A stochastic aggregate model for scheduling ancillary services from V2G-capable EV fleets while considering uncertain driver behaviour has been introduced. Aggregate boundaries for energy and power are used to summarise the potential flexibility of a fleet of EVs. Multiple linear regression is used to forecast the aggregate boundaries, and scenarios are generated based on historic distributions to account for prediction errors. A stochastic model predictive control algorithm is proposed that uses the scenario forecasts for day-ahead scheduling of ancillary services and includes Conditional Value-at-Risk to ensure the algorithm does not make overly optimistic decisions that incur penalties. The proposed algorithm allows aggregators to leverage the size of their EV fleets through aggregation while managing the risks through stochasticity and CVaR.\\
A case study is carried out for reserve service provision in the GB electricity grid. Using a large dataset of domestic EV chargers, an MLR model can predict the boundaries with increasing accuracy, reaching an $R^2$ of 0.7, for a fleet of 1,000, meaning the uncertainty has been reduced by 70\% compared to an algorithm without aggregate predictions. The revenue per EV is observed to increase as the number of EVs in the fleet increases because aggregate driving behaviour becomes more predictable. After reaching a fleet size of around 400 EVs, the increase in revenue per EV levels off. For fleet sizes above 400 EVs an average 1.8kW of reserve power is scheduled, reducing the charging costs by 60\%.\\
The simulation results show that the proposed algorithm allows the grid operator to plan without compromising on consumer needs. So long as consumers are willing to indicate their planned departure, their vehicle battery can reliably provide grid services without any impact on their planned driving. Future work should further investigate charging allocations to individual EVs, including the amount of required energy transferred between EVs. Additionally, the effect of enhanced prediction models could be trialled, and sensitivity analyses for the performance of the algorithm in different market conditions could be carried out, as key elements of future work.

\bibliographystyle{IEEEtran}
\bibliography{bibliography2}

\vfill

\end{document}